\documentclass{aa}
\newcommand{\kms}{km~s$^{-1}$}
\usepackage{graphicx}
\begin{document}
\title{Distances and ages of NGC6397, NGC6752 and 47 Tuc\thanks{Based on
data collected at the European Southern Observatory, Chile, telescopes
(program 165.L-0263)}}

\author{R.G. Gratton\inst{1}, 
A. Bragaglia\inst{2},
E. Carretta\inst{1}, 
G. Clementini\inst{2},
S. Desidera\inst{1},
F. Grundahl\inst{3},
and S. Lucatello\inst{1,4}}

\offprints{R.G. Gratton}

\institute{INAF-Osservatorio Astronomico di Padova, Vicolo dell'Osservatorio
5, 35122 Padova, Italy,\\
\and
INAF-Osservatorio Astronomico di Bologna, Via Ranzani 1, 40127 Bologna,
Italy,\\
\and
Institute of Physics and Astronomy, Aarhus University, Ny Munkegade, 8000 
Aarhus C, Denmark,\\
\and
Dipartimento di Astronomia, Universit\`a di Padova, Italy, Vicolo
dell'Osservatorio 2, 35122 Padova, Italy}

\date{Received: 28-04-03 ; accepted: 23-06-03; version received: 05-06-03}

\abstract{ New improved distances and absolute ages for the Galactic globular
clusters NGC6397, NGC6752, and 47 Tuc are obtained using the Main Sequence
Fitting Method. We derived accurate estimates of reddening and metal abundance
for these three clusters using a strictly differential procedure, where the
Johnson $B-V$\ and Str\"omgren $b-y$ colours and UVES high resolution
spectra of turn-off stars and early subgiants belonging to the clusters were
compared to similar data for field subdwarfs with accurate parallaxes measured
by Hipparcos. The use of a reddening free temperature indicator (the profile
of H$\alpha$) allowed us to reduce the error bars in reddening determinations
to about 0.005 mag, and in metal abundances to 0.04 dex, in the scales defined
by the local subdwarfs. Error bars in distances are then reduced to about 0.07
mag for each cluster, yielding ages with typical random errors of about 1 Gyr.
We find that NGC6397 and NGC6752 have ages of $13.9\pm 1.1$\ and $13.8\pm
1.1$~Gyr respectively, when standard isochrones without microscopic diffusion
are used, while 47 Tuc is probably about 2.6 Gyr younger, in agreement with
results obtained by other techniques sensitive to relative ages. If we use
models that include the effects of sedimentation due to microscopic diffusion
in agreement with our observations of NGC6397, and take into account various 
sources of possible systematic errors with a statistical approach, we conclude
that the age of the oldest globular clusters in the Galaxy is $13.4\pm 0.8\pm
0.6$~Gyr, where the first error bar accounts for random effects, and the
second one for systematic errors. This age estimate is fully compatible with
the very recent results from WMAP, and indicates that the oldest Galactic
globular clusters formed within the first 1.7~Gyr after the Big Bang,
corresponding to a redshift of $z\geq 2.5$, in a standard $\Lambda$CDM model.
The epoch of formation of the (inner halo) globular clusters lasted about
2.6~Gyr, ending at a time corresponding to a redshift of $z\geq 1.3$. On the
other hand, our new age estimate once combined with values of $H_0$\ given by
WMAP and by the HST Key Project, provides a robust upper limit at 95\% level
of confidence of $\Omega_M<0.57$, independently of type Ia SNe, and strongly
supports the need for a dark energy. The new cluster distances lead to new
estimates of the horizontal branch luminosity, that may be used to derive the
zero point of the relation between the horizontal branch absolute magnitude
and metallicity: we obtain $M_V(HB)=(0.22\pm 0.05)({\rm [Fe/H]}+1.5)+(0.56\pm
0.07)$. This zero point is 0.03 mag shorter than obtained by Carretta et al.
(2000) and within the error bar it agrees with, but it is  more precise than
most of the previous individual determinations of the RR Lyrae absolute
magnitude. When combined with the apparent average luminosity of the RR Lyrae
stars in the LMC by Clementini et al. (2003), this zero point provides a new
estimate of the distance modulus to the LMC: $(m-M)_0=18.50\pm 0.09$.
   }

\authorrunning{Gratton R.G. et al.}
\titlerunning{Distances and Ages of Globular Clusters}

   \maketitle
%

\section{Introduction}

Globular clusters are among the oldest objects in our Galaxy. Their ages
provide basic informations on the early stages of Galactic formation, and give
lower limits to the age of the Universe. This latter issue has become in the
last years less urgent, since the increasing evidences for an acceleration in
the Universe expansion have lead to older ages for the Universe. However, when
coupled with determinations of the Hubble constant $H_0$\ from the WMAP
experiment (Spergel et al. 2003) and from the HST Key Project (Freedman et al.
2001), accurate ages of globular clusters can be used to constrain the value of
$\Omega_M$\ independently of observations of type Ia SNe (Perlmutter et
al. 1999). This is an important issue given the existing concerns about
systematic effects on the maximum brightness of type Ia SNe occurring at high
redshift (see e.g. Dominguez et al. 2001). Once the presence of dark
energy is assumed, globular cluster ages can be used to constrain a time
average of the exponent of the equation of state of the vacuum energy $w$\
(Jimenez et al. 2003), possibly allowing to discriminate between various models
(strings, vacuum energy, quintessence: see Wang et al. 2000). On the other
side, assuming a standard $\Lambda$CDM model, accurate ages for globular
clusters and for the Universe (within this framework fixed at $13.7\pm
0.2$~Gyr by the WMAP measurements: Bennett et al. 2003) allow to compare the
epoch of formation of our Galaxy with the evolution of high redshift galaxies,
and to put it into a cosmological framework (see Carretta et al. 2000, and
Freeman \& Bland-Hawthorn, 2002).

Ages require clocks; in the case of globular clusters, three major dating
techniques may be considered. First, nucleocosmochronology provides ages by
comparing abundances of unstable isotopes with those of stable ones. Widely
used in geology, this technique is much more difficult to apply in astronomy
because we may only derive accurate abundances for a few elements, that are
often not tied to each other by simple decaying mechanisms. Assumptions about
nucleosynthesis are needed. Long living unstable elements like Th and more
recently U have been used to derive ages in a few old, very metal-poor stars
by comparing their abundances with those of other stable nuclei synthesized by
the r-process (see Sneden et al. 2001, Cayrel et al. 2001, Cowan et al. 2002).
Recently Hill et al. (2002) obtained an age of $14.0\pm 2.4$~Gyr for the
extremely metal-poor star CS31082-001, employing this technique. In spite of
these admirable efforts, real uncertainties of such determinations are
difficult to assess and are likely large, because they depend on the adopted
model for the r-process that is still quite uncertain (see e.g. Truran et al.
2002).

A second clock relies on the final part of the cooling sequence for white
dwarfs, where it is possible to exploit the well defined maximum of the
luminosity function due to the formation of H$_2$\ molecules in the
atmospheres (Richer et al. 2000). Recently, a spectacular, extremely deep
colour magnitude for M4 obtained using HST has allowed Hansen et al. (2002) to
derive an age of $12.7\pm 0.7$~Gyr for this cluster. Again the error bar is
uncertain, and De Marchi et al. (2002) showed that the maximum of the
luminosity function has not yet been reached, and that the adoption of
different sets of evolutionary models for white dwarfs would lead to quite
different ages. In fact the only solid conclusion that can be reached at
present with this method is that the age of globular clusters is larger than
10~Gyr.

The third clock uses the luminosity of the turn-off (TO) from the main
sequence: this is a well established method, whose uncertainties have long been
studied in detail (see Renzini and Fusi Pecci 1988). Model uncertainties have
been carefully considered by several authors (see e.g. Chaboyer et al. 1996);
the main source of remaining uncertainty is related to the treatment of
microscopic diffusion and levitation due to radiation pressure. These
mechanisms are required to produce solar models that agree with evidences from
helioseismology. Models which include microscopic diffusion (using a complete
ionization approach) lead to ages for globular clusters that are systematically
smaller by about 1~Gyr than those predicted by models where diffusion is
neglected. Other sources of uncertainties (treatment of subatmospheric
convection, details of the used code, and of the normalization to the Solar
values) have a smaller impact (typically less than 0.5~Gyr). Chaboyer et al.
(1996) showed how all these uncertainties can be treated in a statistical way,
leading to a reliable prediction for the error bar.

However, the main source of errors in ages from the TO luminosity actually
comes from errors in the distances (that also affect ages estimated from the
end of the white dwarf cooling sequence, but notably not those from
nucleocosmochronology). Typically, an error of 0.07 mag in distance moduli
(that is, a mere 3.5\% in distances) leads to an error of about 1~Gyr in ages.
Due to the large sensitivity of ages on distances, several investigators have
considered dating techniques that are independent of distance \footnote{For
instance, Chaboyer \& Krauss (2002) have recently derived the age of the
double-lined detached binary OGLE17GC in $\omega$~Cen from the location of the
two components in the colour-magnitude diagram and their masses, independently
of the cluster distance. The derived age t=$11.1\pm 0.7$~Gyr is compatible
with those obtained for the clusters analyzed in the present paper. We also
note that $\omega$~Cen is a very peculiar object, perhaps the nucleus of a now
dissolved nucleated dwarf elliptical, and its age may well be different from
the age of the bulk of Galactic globular clusters.}; however, most of them
only provide relative ages to the clusters that are still very useful in
studies of Galactic formation and evolution. Relative ages for globular
clusters accurate to better than 1~Gyr can be derived mainly with two
techniques: the difference in magnitude between the horizontal branch (HB) at
the average colour of the RR Lyrae stars and the TO (the so-called vertical
method), and the difference in colours between the TO and the base of the
red giant branch (RGB) (the so-called horizontal method). Rosenberg et al.
(1999) showed that the relative rankings obtained with these two techniques
agree well. We will use Rosenberg et al. results in our discussion, that is
however focused on absolute ages.

Derivation of distances to globular clusters at the accuracy needed for their
dating is a difficult task. In perspective, important progresses are expected
in the next few years from the dynamical methods that compare proper motions
obtained using HST with extensive sets of radial velocities obtained from new
generation fiber fed spectrographs like FLAMES, by means of suitable dynamical
models for the clusters (see King \& Anderson 2002; Bedin et al. 2003).
Relevant data will be obtained in the next years, and there is sound hope to
derive robust distance estimates accurate to a few per cent level for several
clusters. In the meantime, we may try to improve other techniques to provide a
comparable accuracy for at least some clusters. In this paper new distances for
three of the four clusters closest to the Sun are derived using the Main
Sequence Fitting Method and data acquired within the ESO Large Program
165.L-0263, at the VLT telescope. In Sect. 2 we discuss the most critical
issues within this method, and our strategy to reduce the 1-$\sigma$\ error
bar from the previous value of $\pm 0.12$~mag (Carretta et al. 2000) to $\pm
0.07$~mag. In Sect. 3 we present the observational data (both photometry and
high resolution spectroscopy) on which our new analysis is based. In Sect. 4
we present our new derivation of reddening, metallicity, and distance to the
clusters.  In Sect. 5 we use our new values of the distances to these three
clusters to estimate the zero point of the luminosity-metallicity relation for
RR Lyrae stars and discuss the impact of our results on the distance to the
Large Magellanic Cloud (LMC). Finally, in Sect. 6 we obtain ages for the
clusters, compare them with previous estimates, and briefly discuss their
impact on the age of the Universe and the epoch of formation of the Milky Way
(MW).

\begin{table}
\caption{Error budget of distance moduli based on the Main Sequence Fitting
Method before and after this paper}
\begin{tabular}{lcc}
\hline
Effect/involved error               & $\Delta(m-M)$ & $\Delta(m-M)$ \\
                                    & Before & After \\
\hline
\\
\multicolumn{3}{c}{Local sample properties}\\
\\
Parallax error                      & $\pm 0.01$ &  $\pm 0.01$ \\
Malmquist bias                      & negligible &  negligible \\
Lutz-Kelker correction              & $\pm 0.02$ &  $\pm 0.02$ \\
\\
\multicolumn{3}{c}{Binary contamination}\\
\\
Field binaries                      & $\pm 0.02$ &  $\pm 0.02$ \\
Cluster binaries                    & $\pm 0.02$ &  $\pm 0.02$ \\
\\
\multicolumn{3}{c}{Systematic differences}\\
\\
Phot. calibrations (0.01 mag) & $\pm 0.04$ &  $\pm 0.04$ \\
Reddening scale (0.015 mag)         & $\pm 0.07$ &  $\pm 0.035$ \\
Metallicity scale (0.1 dex)         & $\pm 0.08$ & (*) \\
\\
Total Uncertainty                   & $\pm 0.12$ & $\pm 0.07$ \\
\hline
\end{tabular}

(*) The effect of an error in the metallicity were included into the error
bar due to reddening (see text)
\label{t:tab1}
\end{table}

\section{The Main Sequence Fitting Method: critical issues}

Prompted by the release of the high precision parallaxes measured by the ESA
Hipparcos satellite (Perryman et al. 1997), numerous determinations of
distances to several Galactic globular clusters using the Main Sequence
Fitting (MSF) method appeared in the the literature in recent years (Reid
1997, 1998; Gratton et al. 1997; Pont et al. 1998, Carretta et al. 2000;
Percival et al. 2002; Grundahl et al. 2002). As a general result all these
studies produced cluster distances distinctly longer by about 15\% (that is
0.3~mag in the distance moduli) than those obtained previously using the same
technique, although with some differences from one author to the other. As
outlined in Gratton et al. (1997) this result is simply because distances to
the local subdwarfs obtained with the Hipparcos parallaxes are larger than
those obtained with the much more uncertain ground based parallaxes.

While the Main Sequence Fitting Method (MSF) is probably the best understood
among the techniques used to derive distances to globular clusters, there are a
number of subtleties that should be carefully considered when accuracies to a
few per cent level are required. In depth discussions of the most relevant
sources of errors were presented in Gratton et al. (1997), Pont et al. (1998),
and Carretta et al. (2000). These last authors published a table listing the
major contributors to the total error budget affecting the MSF distances, that
we have reproduced in the first two columns of Table~\ref{t:tab1}. Errors may
be divided into three separate groups: the first group is more directly
related to the properties of the local subdwarf sample, and includes errors in
individual parallaxes, Malmquist bias, and the Lutz-Kelker correction. The
main contribution by Hipparcos was to reduce these error sources to very small
values, at least for stars and clusters with metallicities larger than
[Fe/H]$=-2$. The second group concerns possible contamination by binaries both
of field stars and globular cluster mean loci. Pont et al. (1998) suggested a
rather large binary correction, however their result was criticized by Gratton
et al. (1997) and Carretta et al. (2000), who concluded that once known or
suspected binaries are eliminated from the sample, the residual correction for
still undetected binaries is likely to be within a few hundredths of a
magnitude.

Finally, the last group of errors listed in Table~\ref{t:tab1} concerns
possible systematic differences between field and cluster stars in various
parameters adopted in the analysis. They generally affect the colours of the
main sequence, and their impact on the distance determinations is due to the
rather steep slope of the main sequence (roughly, $\Delta M_V\sim
5~\Delta(B-V)$) in the colour magnitude diagram. The MSF is influenced by
systematic errors (given in parenthesis in Table~\ref{t:tab1}) in the
photometric calibrations, in the adopted values for the interstellar
reddening, and in the metallicity scales. According to Table~\ref{t:tab1},
these are by far the largest contributors to the total error bar. Could they
possibly cancel out, errors in the MSF distance moduli would be reduced to
0.04 mag, that is an error of only 2\% in distance, and of 0.6~Gyr in age.
While the first two terms in this group (namely photometric calibration and
reddening) directly affect the colours of the stars, the third one
(metallicity) acts indirectly, through the dependence on metallicity of the
main sequence colour in the commonly used bands (mainly $B-V$). Systematic
differences in reddenings may arise from uncertainties in the scale height
of interstellar dust, since local subdwarfs are within the dust absorbing
layer, while globular clusters are much farther. At the same time, systematic
differences in metallicities may be present when comparing abundances obtained
from analysis of local subdwarfs and giants in globular clusters.

In order to significantly reduce the error in the cluster distances derived
with the MSF method, we need then to deal with all these various sources of
errors. We note that all of them could be substantially reduced if we were
able to obtain temperatures and metal abundances for both field and cluster
main sequence stars using the same colour-independent technique, thus
by-passing uncertainties related to colours. This was, in fact, our approach;
however, since the present instrumentation still does not allow to derive
temperatures and colours directly for faint unevolved main sequence stars in
globular clusters, we used instead slightly brighter stars near the TO and at
the base of the RGB. For these stars temperatures were derived by fitting the
profile of H$\alpha$\ in high resolution spectra obtained with UVES at the
Kueyen telescope (ESO VLT Unit 2). Comparisons of the colour-temperature
relations for field (assumed to be unreddened) and cluster stars allowed us to
derive the reddening of the clusters. Actually, these reddening values depend
on the adopted photometries, so that a zero point error in the photometric
calibration directly translates into a corresponding error in the reddening.
However, apparent distances (that is, not corrected for the interstellar
absorption) are unaffected by these errors. The same high resolution spectra
were used to perform an abundance analysis both for field and cluster stars
using strictly the same procedure. In this way, we were able to obtain
reddening and metallicity scales for globular clusters with an accuracy only
limited by the dispersion of individual temperature derivations. It is
unfortunate that this is still not negligible so that the error due to the
reddening scale, while reduced to about half of the original value given in
Table~\ref{t:tab1}, is still the largest individual source of error. Also,
a significant contribution to the error bar is likely due to residual, colour
dependent photometric errors, that do not cancel out using our procedure.
Nevertheless, our distance derivations, with errors of about 3.5\%, are the
most accurate presently available for globular clusters, and allow a
substantial progress in the absolute age derivation.

\begin{table*}
\caption{Unevolved bona fide single local subdwarfs}
\begin{tabular}{lcccccccc}
\hline
HD/DM  &   $\pi$  &  $V$ & $M_V$ & $B-V$ & $b-y$ &$[$Fe/H$]$&$[\alpha$/Fe$]$&$[$M/H$]$\\
       &    mas   &      &       &       &       &          &          &             \\
\hline
\\
\multicolumn{9}{c}{Program stars}\\
\\
$-$35~0360 & $ 16.28\pm 1.76$ &10.25 & $6.31\pm 0.23$ & 0.765 & 0.469 &$-1.15$& 0.39 &$-$0.86 \\
     25329 & $ 54.14\pm 1.08$ & 8.51 & $7.18\pm 0.04$ & 0.870 & 0.529 &$-1.80$& 0.49 &$-$1.43 \\
     75530 & $ 18.78\pm 1.48$ & 9.19 & $5.56\pm 0.17$ & 0.734 & 0.445 &$-0.61$& 0.31 &$-$0.39 \\
$-$80~0328 & $ 16.46\pm 0.99$ &10.10 & $6.18\pm 0.13$ & 0.576 & 0.423 &$-2.03$& 0.17 &$-$1.91 \\
    103095 & $109.22\pm 0.78$ & 6.45 & $6.64\pm 0.02$ & 0.750 & 0.484 &$-1.33$& 0.29 &$-$1.12 \\
    120559 & $ 40.02\pm 1.00$ & 7.97 & $5.98\pm 0.05$ & 0.664 & 0.424 &$-0.94$& 0.33 &$-$0.70 \\
    126681 & $ 19.16\pm 1.44$ & 9.32 & $5.73\pm 0.16$ & 0.597 & 0.400 &$-1.14$& 0.32 &$-$0.91 \\
    134439 & $ 34.14\pm 1.36$ & 9.07 & $6.74\pm 0.09$ & 0.777 & 0.486 &$-1.38$& 0.12 &$-$1.30 \\
    134440 & $ 33.68\pm 1.67$ & 9.44 & $7.08\pm 0.11$ & 0.854 & 0.522 &$-1.45$& 0.19 &$-$1.32 \\
    145417 & $ 72.75\pm 0.82$ & 7.52 & $6.83\pm 0.02$ & 0.820 & 0.505 &$-1.39$& 0.34 &$-$1.15 \\
$+$22~4454 & $ 17.66\pm 1.44$ & 9.50 & $5.73\pm 0.18$ & 0.770 & 0.459 &$-0.60$& 0.30 &$-$0.39 \\
\\
\multicolumn{9}{c}{Additional stars}\\
\\
    104006 & $31.35\pm 1.05$ &  8.89 & $6.38\pm 0.07$ & 0.816 & 0.492 &$-0.81$&(0.29)&$-$0.60 \\
    108564 & $35.30\pm 1.20$ &  9.43 & $7.16\pm 0.07$ & 0.976 & 0.563 &$-0.72$&(0.28)&$-$0.52 \\
    123505 & $20.95\pm 1.65$ &  9.67 & $6.27\pm 0.17$ & 0.782 & 0.475 &$-0.72$&(0.28)&$-$0.52 \\
    145598 & $26.04\pm 1.34$ &  8.65 & $5.73\pm 0.11$ & 0.662 & 0.425 &$-0.79$&(0.28)&$-$0.59 \\
\\
\hline
\end{tabular}
\label{t:tab2b}
\end{table*}

\section{Observational data}

The observational data used in this paper were obtained within the ESO Large
Program 165.L-0263. In the present analysis we consider only three of the
globular clusters observed within that project, namely NGC6397, NGC6752, and
47 Tuc. These clusters were selected because they are the closest to the Sun,
with the only exception of NGC6121 (M4), that is known to be affected by
differential interstellar reddening which greatly complicates its analysis
(see e.g. Ivans et al. 1999). For each of the selected clusters we obtained
high resolution spectra at a resolution of about 40,000 and $20<S/N<100$\ for
stars near the TO, and at the base of the RGB. With the same set up we also
acquired spectra of about thirty local subdwarfs selected from the sample in
Carretta et al. (2000). All of them have parallaxes measured by Hipparcos with
errors $\Delta\pi/\pi<0.12$. Informations about binarity are available for
most of these stars: they were used to select an appropriate sample of bona
fide single stars adopted in the distance derivations. Informations on the
unevolved ($M_V>5.5$)  bona fide single local subdwarf are given in the first
part of Table~\ref{t:tab2b}.

Details of data aquisition and abundance analysis have been given elsewhere
(Gratton et al. 2001, 2003; Carretta et al. 2003), and here we will recall
only a few relevant points.

\subsection{Photometry}

While an enormous amount of high quality photometric data has been obtained in
the last years for globular clusters, most of it could not be used in the
present analysis since these photometries are mainly in the $R$\ and $I$\
photometric bands and we lack comparable data for the local subdwarfs. The
same applies to the excellent HST data. We preferred to avoid uncertain colour
transformations (see Clementini et al. 1999 for a discussion), and throughout
this paper we then used only two sets of photometric data existing both for
field and cluster stars, namely the Johnson $BV$\ and the Str\"omgren $uvby$\
photometry. Johnson $BV$\ photometry for the program clusters was taken from
Alcaino et al. (1997) and Kaluzny (1997) for NGC6397; from Thompson et al.
(1999) for NGC6752; and from Hesser et al. (1987), corrected according to the
prescriptions in Percival et al. (2002), for 47 Tuc. Str\"omgren $uvby$\
photometry was obtained by Grundahl \& Andersen (1999). We will assume
hereinafter that all these photometries are in the standard systems.

Johnson and Str\"omgren photometric data for the field stars were obtained from
the Simbad database: for the $BV$ photometry we simply averaged the various
entries given there, while for the Str\"omgren photometry we adopted the
values recommended by Hauck \& Mermilliod (1998). The adopted photometric data
for the field stars are given in Gratton et al. (2003) and also provided in
Table~\ref{t:tab2b} for the stars used in the distance derivations. Reddening
for the field subdwarfs is discussed by Schuster \& Nissen (1989) and Carney
et al. (1994), and generally found to be small or zero. Hereinafter we will
assume that the field stars are unreddened.

The absolute magnitudes listed in Table~\ref{t:tab2b} do not include the
Lutz-Kelker (1973) correction. This is uncertain, but it is small
for our sample (the maximum value for an individual star being $\sim$0.07 mag).
In our distance derivations, we have corrected for the Lutz-Kelker effect
using Eq. (1) of Gratton et al. (1997), that probably slightly overestimates
this correction. However, cluster distance moduli would only be 0.01 mag
longer we neglect if this correction.

\subsection{High Resolution Spectroscopy}

The high resolution spectra were used to derive effective temperatures from
the profile of H$\alpha$, and the chemical composition from the analysis of
weak metal lines. These analyses are described in Gratton et al. (2001),
Carretta et al. (2003), and Gratton et al. (2003). The same procedure
was adopted both for field and cluster stars.

For the analysis of the H$\alpha$\ line we compared the observed spectral
profiles with those derived using Kurucz (1994) models with the overshooting
option switched off. Synthetic profiles were computed with the same precepts
adopted by Castelli et al. (1997). An example of fit obtained with this
technique is given in Gratton et al. (2003); note that appropriate gravities
and metallicities were adopted when synthesizing line profiles. Temperatures
derived with this approach have typical errors of about 150 K. This rather
large uncertainty is due to difficulties related to a proper flat fielding of
the echelle spectra provided by the 31.6 gr/mm grating that has a rather
limited free spectral range. Fig. 6 of Gratton et al. (2003) compares the
effective temperatures obtained by this technique with those obtained from
colours for the field stars. The r.m.s of the relation is 159 K. Similar
uncertainties are derived from the star-to-star scatter for stars in clusters.
Ten to eighteen stars were considered in each cluster; errors in the zero
points of the temperatures are then from 35 to 50 K for individual clusters,
to be added quadratically to a similar uncertainty for the field stars (about
27 K).

\begin{table*}
\caption{Reddening estimates for the three globular clusters}
\begin{tabular}{lccc}
\hline
Source          &      $E(B-V)$    &      $E(B-V)$    &      $E(B-V)$    \\
                &      NGC6397     &      NGC6752     &      47 Tuc      \\
\hline
Stars used      &        10        &         18       &        12        \\
$b-y$           & $0.178\pm 0.007$ & $0.045\pm 0.007$ & $0.021\pm 0.005$ \\
$B-V$           & $0.186\pm 0.006$ & $0.035\pm 0.007$ & $0.035\pm 0.009$ \\
average         & $0.183\pm 0.005$ & $0.040\pm 0.005$ & $0.024\pm 0.004$ \\
\\
Harris 1996     &       0.18       &       0.04       &       0.05       \\
Schlegel et al. 1998 &       0.187      &       0.056      &       0.032      \\
\hline
\end{tabular}
\label{t:tab2}
\end{table*}

\section{Reddening, metallicity and distances}

\subsection{Reddening}

Estimates of the interstellar reddening toward the selected clusters were
obtained by comparing the observed colour-temperature relations for the field
subdwarfs with that determined for the globular cluster stars. When stars of
similar metallicity and gravity are considered, the reddening is simply given
by the offset between the two relations. However, the field stars generally
have both metallicities and gravities slightly different from those of the
cluster stars, so we inverted the colour-temperature relations given by Kurucz
(1994) model atmospheres both for field and globular cluster stars, using
surface gravities given by the location of the stars in the colour-magnitude
diagram and masses from fittings to 14~Gyr old isochrones (Girardi et al.
2002), and metallicities from our abundance analysis (see Sect. 4.2). We
then compared these colours with those observed for the field and globular
cluster stars: the offsets between colours are our reddening estimates. 

As noticed by the referee, reddenings obtained by this procedure are
sensitive to the selected colour transformations. Somewhat different results
would be obtained by replacing Kurucz transformations with those e.g. from
Houdashelt et al. (2000; we preferred the former ones because they consider
both $B-V$\ and $b-y$\ colours). Replacing Kurucz transformations with those 
from Houdashelt et al., and repeating the whole procedure, we find that the
reddening from $B-V$\ reduces by as much as 0.02 mag for NGC6397, while
changes are about half this value for NGC6752, and very small for 47 Tuc. We
expect that uncertainties in the transformations are much less important for
the intermediate band $b-y$\ colour than for the wide band $B-V$\ one. However,
this cannot be verified since Houdashelt et al. (2000) do not provide
predictions for $b-y$. 

Three things must be noticed here: (i) a strong support to the transformations
we have adopted is provided by the very good agreement between the temperature
scale adopted in our analysis and that obtained using the Infrared Flux Method
by Alonso et al. (1996; see Gratton et al. 2003: note that this excellent
agreement is obtained for the average of temperatures from $B-V$\ and $b-y$\
colours). (ii) Both cluster subgiants and TO-stars were used, and results
averaged: typically, TO-stars have temperatures of about 5800-6300 K, and
subgiants of about 5100-5500~K, depending on the cluster. (iii) The field
stars used in the cluster reddening derivation are those listed in Gratton et
al. (2003) having temperatures from H$\alpha$. A total of 35 stars
were used; they have 4900$<T_{\rm eff}<$6300~K, 3.4$<\log g<$4.8, and
$-2.2<[M/H]<-0.4$, so that they effectively overlap with the cluster stars,
reducing the impact of offsets due to uncertainties in the temperature-colour
transformations used as intermediate steps in our reddening derivations. This
is confirmed by the good agreement between the reddening values obtained from
subgiants and TO-stars: on average, the difference for the tree clusters is
$0.006\pm 0.005$~mag (reddenings from TO-stars being slightly smaller), and in
no case the difference between the values obtained from TO-stars and subgiants
is larger than the 1~$\sigma$\ internal error bar.

As mentioned above, the procedure was repeated twice, both using the Johnson
$B-V$\ and the Str\"omgren $b-y$\ colours. Corresponding reddening estimates
are provided in Table~\ref{t:tab2} in units of $E(B-V)$. Values derived from
Str\"omgren $b-y$\ colours were converted to $E(B-V)$ using the transformation
$E(B-V)=E(b-y)/0.72$\ (Crawford \& Mandwewala 1976). The final reddening
adopted for each cluster is the weighted average of the values obtained from
the two colours, where errors in the values for each colour are obtained from
the dispersion of individual data. These reddenings are $E(B-V)=0.183\pm
0.005$\ for NGC6397, $E(B-V)=0.040\pm 0.005$\ for NGC6752, and
$E(B-V)=0.024\pm 0.004$\ for 47 Tuc. Systematic uncertainties due to possible
offsets between the temperature scales for field and cluster stars caused by
errors in individual temperature determinations were evaluated as follows. In
total, 40 globular cluster and 35 field stars were used (these include field
stars brighter and more evolved than those listed in Table~\ref{t:tab2b}).
Adopting a common error of $\pm 150$~K for individual temperatures from
H$\alpha$, the possible offset between the two sets is of $\pm 36$~K from
simple statistics. This yields zero point errors of $\pm 0.011$~mag in the
reddening (producing an error of $\pm 0.055$~mag in the distance) and of $\pm
0.025$~dex in the metal abundance (that in turn transforms into a typical
error of $\pm 0.020$~mag in distance). However, it should be noted that these
two changes in distance are of opposite sign, so that the systematic
uncertainty in distance moduli becomes $\pm 0.035$~mag.

The last two rows of Table~\ref{t:tab2} list literature reddening
determinations taken from the compilation by Harris (1996) and from the maps
of interstellar reddening obtained by Schlegel et al. (1998) from analysis of
the COBE-DIRBE data. The agreement between the various determinations is
excellent, in particular with the latter ones: our values are on average
$0.008\pm 0.012$\ smaller than those listed by Harris, and $0.009\pm 0.005$~mag
smaller than Schlegel et al.'s. We note that our reddening values might be
systematically smaller than those derived by other authors because we have
assumed that the local subdwarfs are unreddened; however, only three of the
35 field stars considered here were found to be reddened by Schuster \&
Nissen (1989) and Carney et al. (1994) (namely, HD116064, HD132475, and
HD140283); the average reddening of the whole sample is then
$E(B-V)=0.003$~mag. Once this is taken into account, the small residual
differences are well within the uncertainties of the various determinations.
We in fact estimate that the error on the zero points of our reddenings is 
$\pm 0.011$~mag from the uncertainty in the zero-point of the relative
temperature scales. We neglect here the possible impact of the adopted 
colour-temperature calibration, that we cannot estimate lacking adequate data
for $b-y$~colours. The agreement with these independent estimates supports
the small error bars we attach to our reddening estimates.

\begin{figure} 
\includegraphics[width=8.8cm]{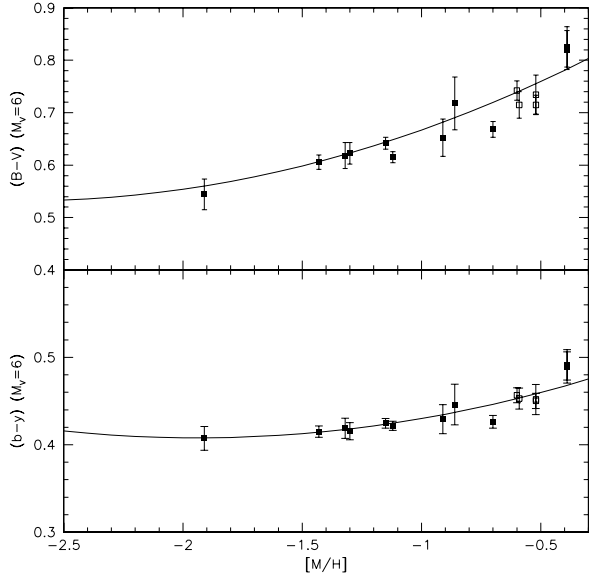}
\caption[]{Colour of the main sequence at an absolute magnitude $M_V=6$. The
upper panel is for the Johnson $BV$\ photometry, the lower panel is for the
Str\"omgren $by$\ photometry. Symbols represent individual unevolved
subdwarfs (that is, stars with $M_V>5.5$) with parallax error
$\Delta\pi/\pi<0.12$. The lines are the model predictions by Straniero et al.
(1997). Filled symbols are the field stars considered in this paper; open
squares are additional unevolved subdwarfs, used as a sanity check (see text)}
\label{f:fig1}
\end{figure}

\subsection{Metallicity}

Metal abundances (both [Fe/H] values and average overabundance of the
$\alpha-$elements [$\alpha$/Fe]) were derived both for field and clusters stars
using strictly the same procedure. The observed field stars cover the metal
abundance range from [Fe/H]=$-$0.6 to $-$2.0. Within this scale, the three
clusters have metal abundances of [Fe/H]=$-2.03\pm 0.05$, [Fe/H]=$-1.43\pm
0.04$, and [Fe/H]=$-0.66\pm 0.04$ for NGC6397, NGC6752 and 47 Tuc,
respectively. These cluster abundances are quite similar to those obtained by
Carretta \& Gratton (1997). It must be noticed that the metal abundances
derived for the field stars in the present analysis are lower than those used
by Carretta et al. (2000). Discarding HD145417, the average difference is
$0.13\pm 0.04$~dex. This difference is due to the lower temperatures adopted
in the present analysis. This is the first time that metal abundances
consistently derived for field and cluster stars in similar evolutionary
phases are used when deriving distances with the Main Sequence Fitting Method.

Fig.~\ref{f:fig1} displays the run of the $B-V$\ and $b-y$\ colours with
metallicity for the field stars in Table~\ref{t:tab2b} at an absolute magnitude
$M_V=6$. In this figure we have used the overall metallicity [M/H]. This is
connected to [Fe/H] by the relation [M/H]=[Fe/H]+$\log (0.638 f + 0.362)$,
where $f$\ is the average overabundance of the $\alpha-$elements (Mg, Si, Ca,
and Ti) by number (not in the logarithm; see Straniero et al. 1997). Note that
$\log f=0.34$, 0.29, and 0.30 for NGC6397, NGC6752, and 47 Tuc respectively
(Gratton et al. 2001; Carretta et al. 2003). Appropriate values for the field
stars are listed in Table~\ref{t:tab2b}. Following Gratton et al. (1997), this
comparison was made by correcting the observed colours for the difference due
to the absolute magnitude of the stars, by moving the stars parallel to the
main sequence. The adopted corrections are $\Delta
(B-V)=-(M_V-6)[0.192+0.028~(M_V-6)]$, and $\Delta (b-y)=-0.097~(M_V-6)$. The
lines overimposed to the data in Fig.~\ref{f:fig1} are not best fit lines,
but simply the predictions of the theoretical isochrones by Straniero et al.
(1997; as discussed in Gratton et al. 1997 and Carretta et al. 2000, other
isochrone-sets give similar predictions once the same transformations from
the theoretical to observational planes are considered). They are represented
by the following equations:
\begin{equation}
(B-V)_{(M_V=6)} = 0.876 + 0.257 [M/H] + 0.048 [M/H]^2,
\end{equation}
and:
\begin{equation}
(b-y)_{(M_V=6)} = 0.502 + 0.097 [M/H] + 0.025 [M/H]^2.
\end{equation}

The agreement between theoretical predictions and observations is excellent,
although a small offset is present in the Johnson $B-V$\ colour. On average,
stars are bluer than predicted by models, by $0.014\pm 0.006$~mag in $B-V$\
and $0.002\pm 0.003$~mag in $b-y$. These small differences may be ascribed to
small uncertainties in the zero point of the photometric pass bands, as well as
to an error in the temperature scale of about 30~K. However they have no
impact on our distance and age derivation, which only use these equations in a
differential way. Once these small offsets are taken into account, the
residuals from the theoretical curves agree quite well with observational
errors in the parallaxes (the dominant contributor), in the colours, and in
the metal abundances: in fact, the reduced $\chi^2$\ values are 1.53 for
$B-V$\ and 0.72 for $b-y$. The only star showing a slightly discrepant value
is star HD 120559 ([M/H]=$-0.70$), that is slightly bluer than expected for
its metallicity.

The scatter around the mean relations is rather large for the metal-rich
stars. Also our subdwarf sample contains only very few metal-rich stars, thus
raising doubts on whether the theoretical relations well predict colours for
[Fe/H]$>-0.6$. As a sanity check of the colour-metallicity relations at high
metal abundances we have considered additional unevolved subdwarfs (i.e. stars
with $M_V>5.5$) taken from the papers of Grundahl et al. (2002) and Percival
et al. (2002). There are four stars satisfying our selection criteria
(unreddened, bona fide single stars, with complete photometric data and
accurate parallaxes). These stars are listed in the lower part of
Table~\ref{t:tab2b}. Since we do not have spectra for these stars, their metal
abundances were obtained by correcting the values listed by Grundahl et al.
(2002) for the average offset between the metallicities listed in that paper
and the [Fe/H] values obtained in our analysis (Gratton et al. 2003) for stars
in common between the two samples: this is $0.06\pm 0.02$~dex, with our [Fe/H]
values being smaller. Furthermore, we adopted for the [$\alpha$/Fe] values the
averages obtained for stars of the appropriate metallicity by Gratton et al.
(2003). In this way, colours and metallicities for these stars should be
perfectly consistent with those of our program field stars. Data for the
additional stars were plotted as open squares in our figures. They agree
perfectly with the average relation found for our sample: on average they are
bluer than the theoretical predictions by $0.025\pm 0.008$~mag in $B-V$, and
by $0.004\pm 0.003$~mag in $b-y$. Within the errors, these are the same
offsets found for the program stars.

The agreement between theoretical predictions and observations is a strong
support to the use of colour corrections based on theoretical models when
comparing stars and clusters with different metallicities.

Finally we note that the Str\"omgren $b-y$\ colour has a much weaker
dependence on metallicity than the Johnson $B-V$\ colour, and appears then
definitely superior in the analysis of the most metal-poor clusters, for which
some extrapolation off the sequence of local subdwarfs is needed.

\begin{figure*} 
\includegraphics[width=13cm]{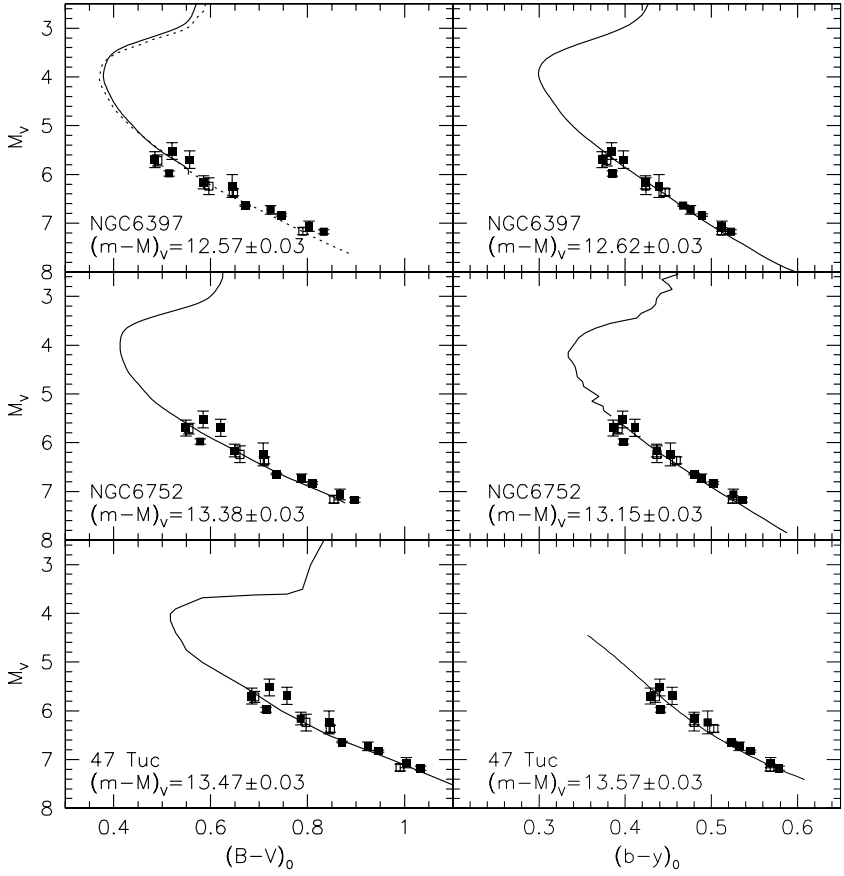}
\caption[]{Main Sequence Fitting distances to the program clusters (NGC6397:
upper panels; NGC6752: middle panels; 47 Tuc: lower panels). For each cluster
fits obtained both from Johnson $B-V$\ (left) and Str\"omgren $b-y$\
(right) colours are shown. The parameters adopted in these fits are listed in
Tables~2 and 3. Data for individual bona fide single subdwarfs are shown with
their error bars. Solid and dashed lines in the upper left panel are the
cluster mean loci from Kaluzny (1997) and Alcaino et al. (1997), respectively.
Filled symbols are the subdwarfs used to estimate the distance moduli (upper
portion of Table~\ref{t:tab2b}); open squares are additional unevolved
subdwarfs, used as a sanity check (see text)
}
\label{f:fig2}
\end{figure*}

\subsection{Distances}

Distances to each cluster were obtained by fitting the observed globular
cluster mean loci to the location of the local subdwarfs in the colour
magnitude diagram. Only the bona fide single subdwarfs with absolute magnitude
$M_V>5.5$\ listed in the upper portion of Table~\ref{t:tab2b} were used to fit
each cluster locus, to avoid objects possibly evolved off the main sequence.
Colours for each field star were corrected for the difference in metallicity
between the star and the clusters, using the above theoretical relations (1)
and (2), from Straniero et al. (1997) isochrones. Distances obtained from each
individual subdwarf were averaged weighting them according to their errors.
These were the sum in quadrature of the uncertainties in the absolute
magnitude due to errors in the parallaxes, and of the uncertainties due to
errors in colours (assumed to be 0.01 mag in $B-V$, 0.007~mag in $b-y$) and
metallicity (assumed to be $\pm 0.04$~dex).

Derived distance moduli were then shortened by 0.02 mag to account for the
possible presence of undetected binaries; this small correction is the same
adopted by Carretta et al. (2000). Relevant data and results are given in
Table~\ref{t:tab3}.

\begin{table*}
\caption{Distances and ages for globular clusters}
\begin{tabular}{lccc}
\hline
Parameter      &     NGC6397     &      NGC6752    &      47 Tuc     \\
\hline
$[$Fe/H$]$     & $-2.03\pm 0.05$ & $-1.43\pm 0.04$ & $-0.66\pm 0.04$ \\
$[\alpha$/Fe$]$& $~0.34\pm 0.02$ & $~0.29\pm 0.02$ & $~0.30\pm 0.02$ \\
$[$M/H$]$      & $-1.79\pm 0.04$ & $-1.22\pm 0.04$ & $-0.45\pm 0.04$ \\
\\
$(m-M)_{V(B-V)}$&$12.57\pm 0.03$ & $13.38\pm 0.03$ & $13.47\pm 0.03$ \\
$(m-M)_{V(b-y)}$&$12.62\pm 0.03$ & $13.14\pm 0.03$ & $13.57\pm 0.03$ \\
$<(m-M)_V>$    & $12.60\pm 0.08$ & $13.26\pm 0.08$ & $13.52\pm 0.08$ \\
bin. corrected & $12.58\pm 0.08$ & $13.24\pm 0.08$ & $13.50\pm 0.08$ \\
\\
$V$(TO)        & $16.56\pm 0.02$ & $17.39\pm 0.03$ & $17.68\pm 0.05$ \\
$V$(HB)        & $13.11\pm 0.10$ & $13.84\pm 0.10$ & $14.13\pm 0.10$ \\
$M_V$(TO)      &  $3.98\pm 0.08$ &  $4.15\pm 0.08$ &  $4.18\pm 0.08$ \\
$M_V$(HB)      &  $0.53\pm 0.13$ &  $0.60\pm 0.13$ &  $0.63\pm 0.13$ \\
\\
Age (no diff.)~(Gyr) &  $13.9\pm 1.1$  &  $13.8\pm 1.1$ &   $11.2\pm 1.1$  \\
Age (diff.)~(Gyr)    &  $13.5\pm 1.1$  &  $13.4\pm 1.1$ &   $10.8\pm 1.1$  \\
\hline
\end{tabular}
\label{t:tab3}
\end{table*}

Two different estimates of the distance to each cluster were obtained in this
way, corresponding to fitting the $V-(B-V)$\ diagram in the Johnson system,
and the $V-(b-y)$\ diagram in the Str\"omgren system, respectively. The
attached error is the fitting error produced by the individual star errors.
Fig.~\ref{f:fig2} displays the fits we obtained for the three clusters in
each colour. Adopted distances are the simple averages of the values obtained
with each colour. There is no significant systematic difference between
distances obtained from Johnson and Str\"omgren photometry, the average
difference being $0.03\pm 0.12$ mag. The errors expected from uncertainties in
data for individual stars are actually much smaller (about 0.025~mag).
The worst case is that of NGC6752: in this case the distance moduli
given by $B-V$\ is 0.24~mag larger than that given by $b-y$, a difference
much larger than the internal errors of the fits. Adopted metallicities and
reddenings are unlikely to be the cause for this difference - e.g. a
metallicity as low as [M/H]=$-1.8$\ (corresponding to [Fe/H$\sim -2$) is
required to cancel it out; an error of 0.1~dex (yielding a metal abundance
coincident with that estimated by Kraft \& Ivans, 2002, for this cluster) may
only justify a difference of about 0.04~mag. On the other side, a part of the
difference is due to the use of a unique reddening value in both
determinations: if we had rather used the reddening estimates obtained
separately for each colour (thus taking into account possible zero-point
offsets in the colour calibrations), the difference would have been reduced
to 0.18~mag. We think that the residual difference is mainly due to
inaccuracies in the colour term of the photometric transformations from one
system to the other. This is also supported by some small mismatches between
the globular cluster mean loci and the location of the local subdwarfs,
visible in various panels of Fig.~\ref{f:fig2}. These have impact on the
distance derivations because the stars used to derive distances are redder
than those used to derive reddening and metal abundances. However, we cannot
judge from our photometry alone which of the two results - from $B-V$\ or
from $b-y$, is correct. Luckily, differences are much smaller for the two
other clusters. Averaging data for the three clusters, we then estimated that
the internal errors in the distance moduli are $\pm 0.074$~mag.

We note that distances do not change if the four additional stars
used as sanity check (see Sect.~4.2) are used in the fittings.

Systematic errors in these distance estimates are mainly due to the zero point
of our effective temperatures; as discussed in Sect. 4.1 they are of about
$\pm 0.04$ mag. The final error bars of our distance moduli were obtained by
summing quadratically these two contributions.

A small systematic error in our distances is actually related to
uncertainties in the metallicity scale, due to the quadratic form of
the metallicity-colour relations. Our metallicities for these three clusters
are very similar to those obtained very recently by Kraft \& Ivans (2002),
using analysis of giant stars with different model atmospheres: our [Fe/H] 
values are $0.05\pm 0.06$~dex higher than those obtained by them using the
MARCS model atmospheres, and $0.01\pm 0.04$~dex lower than those obtained with
the Kurucz model atmospheres (using Fe~II lines, their preferred abundance
estimators). However, Kraft \& Ivans conclude that various sources of possible
errors make this scale uncertain by at least 0.03-0.05~dex. A similar result
is obtained by comparing our abundances for the field stars with other
recent studies (see Table 10 of Gratton et al. 2003). To estimate the
impact of the uncertainty in the metallicity scale on our distance
derivations, we repeated them by arbitrarily increasing the metallicities for
both field and cluster stars by 0.1 dex (likely, an estimate of the maximum
zero point errors in our metallicities). We found that the distance modulus
to NGC6397 would be reduced by 0.04 mag, that to NGC6752 by 0.01 mag, and
finally that to 47 Tucanae would be increased by 0.03 mag. While not
negligible, this source of errors appears to be much smaller than that due to
the reddening and photometric errors.

\subsection{Comparison with previous determinations}

The literature on distance determinations to the program clusters is very
extensive and here we only consider a few results. Table~\ref{t:tab4} compares
the present distances with previous estimates obtained using the MSF
technique. On average, our new distances are shorter: differences are $0.16\pm
0.06$ mag and $0.08\pm 0.01$\ with Reid (1998) and Carretta et al. (2000),
respectively\footnote{A downward correction to MSF distances by about 0.1 mag
was already suggested by Carretta et al. (2000) based on a comparison of
distances derived by this technique with other values. This difference was
there attributed to small inconsistencies in the reddening and metallicity
scales for globular cluster and local subdwarfs. This correction is confirmed
both in sign and size by the present analysis that uses a new homogeneous
evaluation of reddenings and metallicities for globular clusters. Indeed, our
distance scale is very similar to that final one adopted by Carretta et al.
(2000)}. While the source of parallaxes for the local subdwarfs in all these
various analyses is always the Hipparcos catalogue, a difference of our study
is that we used both the Johnson and Str\"omgren photometry, and we also made
different assumptions about metallicities and reddenings for both cluster and
field stars (values adopted for the clusters in each analysis are listed in
the second and third columns of Table~\ref{t:tab4}).

Differences between the various distance estimates are large for 47 Tuc, with
values spanning a range of 0.35 mag (our new determination is in the middle of
the other values). To understand these differences, first we notice that Reid
(1998) and Carretta et al. (2000) used the Hesser et al. (1987) $BV$\
photometry  without any correction, while we have corrected the 47 Tuc
photometry according to Percival et al. (2002). All other terms held constant,
this causes the distance modulus of this cluster to be 0.19 mag smaller.
Furthermore, while the cluster Fe abundance adopted in the various analyses
appears to be quite similar, there are significant differences in the
abundances adopted for the comparison field stars. Reid (1998), Carretta et
al. (2000), and Percival et al. (2002) all adopt the same metal abundances,
essentially consistent with the scale defined by Carretta et al. (2000).
However, as we already mentioned, this is 0.13 dex higher than that adopted
here, that is obtained from an analysis which is fully consistent for field
and cluster stars. At the metallicity of 47 Tuc, this corresponds to a
difference of 0.25 mag in the distance modulus, with the present one being
longer. Finally, the reddening adopted here (for the first time, consistently
derived for field and cluster stars) is 0.031 mag smaller than that adopted by
Carretta et al. (2000) and Percival et al. (2002), and 0.016 mag smaller than
that considered by Reid (1998). This causes our distance modulus to be 0.15
mag shorter than that in Carretta et al. (2000) and Percival et al. (2002),
and 0.08 mag shorter than that in Reid (1998). When comparing our results with
those of Grundahl et al. (2002), it should be first noticed that different
colours are used (we use the $b-y$\ colour, while Grundahl et al. use the
$v-y$\ colour). Furthermore, there is a difference in the adopted reddening,
and again in the metal abundance of the field stars (their finally adopted
values being larger than ours by 0.16 dex, because they corrected upward by
0.1 dex the abundances originally obtained for the field stars in order to
reach a claimed consistency with the value obtained for the cluster).
 
These different assumptions about photometry, metal abundances and reddenings
fully explain the differences between our new determinations and the above
literature values: they also emphasize the importance of consistent
derivations of metal abundances and reddenings for field and cluster stars.

Differences are smaller for NGC6752; they again may be explained by the
different assumptions about metal abundances and reddenings. In the case
of NGC6397 (not included in the sample considered by Carretta et al. 2000), we
have used a metallicity $\sim$0.2 dex lower (this causes a 0.12 mag shortening
of the distance modulus), and a reddening about 0.007 mag smaller (that
translates into a distance modulus 0.04 mag shorter) than those adopted by
Reid (1998), and this explains about two thirds of the 0.25 mag difference
found in the distance moduli for this cluster.

\begin{table}
\caption{Comparison with previous distance determinations using the Main
Sequence Fitting Method based on Hipparcos parallaxes for the local subdwarfs.
Errors quoted are those given in the original papers; they have different
meanings from the present error bar}
\begin{tabular}{lccc}
\hline
Author         & $[$Fe/H$]$ & $E(B-V)$ & $(m-M)_V$ \\
\hline
\multicolumn{4}{c}{NGC 6397}\\
Reid 1998            & $-1.82$ & 0.19  & $12.83\pm 0.15$ \\
This paper           & $-2.03$ & 0.183 & $12.58\pm 0.08$ \\
\multicolumn{4}{c}{NGC 6752}\\
Reid 1998            & $-1.42$ & 0.04  & $13.28\pm 0.15$ \\
Carretta et al. 2000 & $-1.43$ & 0.035 & $13.32\pm 0.04$ \\
This paper           & $-1.43$ & 0.040 & $13.24\pm 0.08$ \\
\multicolumn{4}{c}{47 Tuc}  \\
Reid 1998            & $-0.70$ & 0.04  & $13.68\pm 0.15$ \\
Carretta et al. 2000 & $-0.67$ & 0.055 & $13.57\pm 0.09$ \\
Percival et al. 2002 & $-0.67$ & 0.055 & $13.37\pm 0.11$ \\
Grundahl et al. 2002 & $-0.67$ & 0.04  & $13.33\pm 0.10$ \\
This paper           & $-0.66$ & 0.024 & $13.50\pm 0.08$ \\
\hline
\end{tabular}
\label{t:tab4}
\end{table}

\begin{table}
\caption{Comparison between our distance determinations and those from
the White Dwarf cooling sequence}
\begin{tabular}{lcc}
\hline
Author          & $E(B-V)$ & $(m-M)_V$ \\
\hline
\multicolumn{3}{c}{NGC 6752}\\
Renzini et al. 1996  & 0.04  & $13.17\pm 0.13$ \\
This paper           & 0.040 & $13.24\pm 0.08$ \\
\multicolumn{3}{c}{47 Tuc}  \\
Zoccali et al. 2001  & 0.055 & $13.27\pm 0.13$ \\
This paper           & 0.024 & $13.50\pm 0.08$ \\
\hline
\end{tabular}
\label{t:tab5}
\end{table}

Among the various other distance determinations, we considered only those
obtained from fitting the white dwarf cooling sequence observed in globular
clusters with a sequence of local white dwarfs with accurate parallaxes. This
comparison is shown in Table~\ref{t:tab5}. Distances from the white dwarf
cooling sequence are independent of metallicity (at least as long as the
assumption that the white dwarf masses do not depend on metallicity is
correct); however they have a dependence on interstellar reddening quite
similar to that obtained for the MSF method, because the two sequences are
roughly parallel in the colour-magnitude diagram. We found good agreement
with the results of Renzini et al. (1996) for NGC6752, while the agreement is
far from being satisfactory for 47 Tuc, whose distance modulus from the white
dwarf cooling sequence (Zoccali et al. 2001) is 0.23 mag shorter than our MSF
value (see last column of Table~\ref{t:tab5}). Indeed the difference is even
larger: in fact, if the same reddening values were adopted, the offset would
be as large as 0.38 mag. The reason for such a large discrepancy between these
two distance determinations for this cluster remains unexplained.

\section{Luminosity of the horizontal branch}

\subsection{Derivation of the absolute luminosity of the HB}

Once the distance of a cluster is known, derivation of the absolute luminosity
of the HB, $M_V{\rm (HB)}$, from the observed HB apparent luminosity, $V{\rm
(HB)}$, seems quite straightforward. $M_V{\rm (HB)}$ is usually defined as the
$V$ absolute magnitude of an HB star at $\log T_{\rm eff}$=3.85, that is taken
to represent the equilibrium temperature of an RR Lyrae star near the middle of
the instability strip. However, since the HB is not exactly horizontal, not
even in the $V$\ band, and since the three clusters considered in the present
paper lack a significant population of RR Lyrae stars, the estimate of their
HBs average luminosities is not so easy.  Furthermore, care should be devoted
to ensure that photometry of both HB and main sequence stars in the cluster
are on the same scales. To ensure proper handling of these issues, we adopted
simply the magnitude difference between TO and HB measured for the three
clusters by Rosenberg et al. (1999), who carefully considered these points. By
combining Rosenberg et al.'s values with our TO absolute magnitudes (see
Sect. 6.1), we obtained HB absolute magnitudes of $0.53\pm 0.13$, $0.60\pm
0.13$, and $0.63\pm 0.13$\ for NGC6397, NGC6752, and 47 Tuc, respectively (see
Table~\ref{t:tab3}). The error bars attached to these $M_V{\rm (HB)}$'s were
obtained by combining those of Rosenberg et al. (1999), with the errors in our
estimates of the absolute magnitude of the TO.

The three values may be averaged together by adopting a relation between the
absolute magnitude of the HB and metallicity. This is generally described in
terms of a linear relation. However, claims have been made that the relation
is not {\it strictly linear} (Caputo et al. 2000, Rey et al. 2000, and
references therein) and doubts have been cast on whether a {\it universal}
luminosity-metallicity relations holds for HB stars in different environments,
e.g. the MW field versus cluster stars, and the MW versus other Local Group
galaxies. Indeed, even at fixed metallicity there is an intrinsic spread in
the HB luminosity due to evolutionary effects whose extent varies as a
function of metal abundance (Sandage 1990), and it has been shown that the
luminosity-metallicity relation may not be strictly linear because it depends
also on the HB morphology and stellar population (Caputo et al. 2000, Demarque
et al. 2000). In principle, since our three clusters span a range in
metallicity of about 1.4 dex, their $M_V({\rm HB})$ values could be used to
derive the slope of the luminosity-metallicity relation for Galactic globular
clusters. However, because of the very small number of clusters and the still
rather large error bars of the $M_V({\rm HB})$ values, the result is very
uncertain: $\Delta M_V({\rm HB})$/$\Delta$[Fe/H]=0.07$\pm$ 0.13 mag
dex$^{-1}$. The definition of the slope of the luminosity-metallicity
relation for HB stars is beyond the purposes of the present paper; to correct
the $M_V({\rm HB})$ values of our three clusters to a common metallicity we
use literature results. A quite accurate and robust estimate for the slope
has been recently obtained by Clementini et al. (2003), based on observations
of a hundred RR Lyrae stars in the LMC. The slope obtained in that paper
($0.21\pm 0.05$ mag dex$^{-1}$) is in very good agreement with that been
determined by Rich et al. (2002; 0.22 mag dex$^{-1}$) from the mean HB
magnitudes at the middle of the instability strip of 19 globular clusters in
M31; it also agrees with the slope found by the Baade-Wesselink analysis of
the MW field RR Lyrae stars (0.20$\pm 0.04$ mag dex$^{-1}$: Fernley et al.
1998). Theoretical models of helium-burning stars also predict the slope of
the luminosity metallicity relation for HB stars\footnote{Actually,
theoretical models predict the luminosity of the Zero Age Horizontal Branch
(ZAHB) and its dependence on metallicity. The correction of the theoretical
$M_V({\rm ZAHB})$-[Fe/H] relation to the observed $M_V({\rm HB})$-[Fe/H]
relation is made either applying a fixed  offset that takes into account
evolution (generally of the order of 0.08-0.10 mag) or using an empirical
correction as the one derived by Sandage (1993) $\Delta V({\rm ZAHB - HB}) =
0.05 {\rm [Fe/H]} + 0.16$. This relation corresponds to an evolutionary
correction of about 0.085 mag at [Fe/H]=$-1.5$.}. Several uncertainties affect
the theoretical models, that explain the scatter observed among different
families of HB models. However, all the most recent models (Caloi et al. 1997,
Cassisi et al. 1999, Ferraro et al. 1999, Demarque et al. 2000, VandenBerg et
al. 2000) agree that although the relation is not universal and not strictly
linear, it can be roughly described by a linear relation with average slope
$\sim$0.23, as a first approximation. We adopted the value $\Delta M_V({\rm
HB})$/$\Delta$[Fe/H]=0.22~mag~dex$^{-1}$ as the average of the above
independent estimates and attached to this slope a conservative error of
$\pm$0.05 mag dex$^{-1}$ (Clementini et al. 2003). We derived
$M_V(HB)=0.65\pm 0.13$, $0.59\pm 0.13$, and $0.45\pm 0.13$\ for NGC6397,
NGC6752, and 47 Tuc respectively, giving a zero point of $0.56\pm 0.07$ at
[Fe/H]=$-1.5$. The error bar is obtained combining the internal errors of the
values obtained for each cluster. The relation between absolute magnitude of
the HB and metallicity adopted from our analysis of the three clusters is
then:
\begin{equation}
M_V(HB) = (0.22\pm 0.05) {\rm ([Fe/H] + 1.5)} + (0.56\pm 0.07).
\end{equation}

\subsection{Comparison with other determinations and implications: the
distance to the LMC}

The zero point of the absolute magnitude-metallicity relation has been
determined by a large variety of methods and approaches; a recent,
comprehensive review is provided by Cacciari \& Clementini (2003). In their
Table 2, that we reproduce in our Table~\ref{t:tab7} for ease of comparison,
these authors provide a summary of determinations of $M_V{\rm (RR)}$ at
[Fe/H]=$-1.5$ obtained by several independent methods. Our zero point in Eq.
(3) (see last row of Table~\ref{t:tab7}) agrees with the results from the
trigonometric parallax of RR Lyrae itself $M_V{\rm (RR)}$=0.61$\pm$0.11
(Benedict et al. 2002), at [Fe/H]=$-1.4$ (Clementini et al. 1995), and with
Koen \& Laney (1998) re-evaluation of Gratton (1998) trigonometric parallaxes
for several HB stars: $M_V{\rm (RR)}$=0.62$\pm$0.11 at [Fe/H]=$-1.5$. Very
good agreement also exists with Cacciari et al. (2000) revision of the
Baade-Wesselink absolute magnitude of the RR Lyrae  star RR Ceti: $M_V{\rm
(RR)}=0.55 \pm 0.12$ at [Fe/H]=$-1.5$, and with the prediction both of the
pulsation models and of the Fourier parameters (see Table~\ref{t:tab7} and
discussion in Cacciari \& Clementini 2003). Concerning the theoretical
models, our $M_V{\rm (HB)}$\ is in excellent agreement with the family of
models that predict a faint value of $M_V{\rm (RR)}$=0.56$\pm$0.12, namely
Ferraro et al. (1999), Straniero et al. (1997), Demarque et al. (2000), and 
VandenBerg et al. (2000) models. Agreement is less satisfactory, but still
within the error bars, with the models by Caloi et al. (1997) and Cassisi et
al. (1999) who predict a brighter zero point of $M_V{\rm
(RR)}$=0.43$\pm$0.12. The statistical parallaxes is the only technique that
disagrees and gives a value about 0.2 mag fainter than the present $M_V{\rm
(HB)}$. Cacciari and Clementini (2003) discuss in detail the problems
connected to this technique and conclude that a more accurate and detailed
modeling of the stellar motions in the Galaxy and larger sample of stars are
needed to provide reliable and robust results.

Clementini et al. (2003) have recently derived a very accurate estimate of the
average luminosity of the RR Lyrae variables in the bar of the LMC: $V_{0}{\rm
(RR)}$=19.064$\pm$0.064 at [Fe/H]=$-1.5$. Combined with our new estimate of
the absolute luminosity of the HB, this leads to a distance modulus to the LMC
of $(m-M)_{0}$=18.50 $\pm$0.09 mag, in perfect agreement with the value
obtained by Clementini et al. (2003).

\begin{table}
\caption{Comparison between various determinations of the luminosity of the
horizontal branch (from Cacciari \& Clementini 2003)}
\begin{tabular}{lc}
\hline
Method& $M_{V}(RR)$\\
      & at [Fe/H]=$-1.5$\\
\hline
Statistical parallaxes& 0.78$\pm$0.12\\ 
Trigonometric parallaxes (RR Lyr)& 0.58$\pm$0.13\\ 
Trigonometric parallaxes (HB stars)& 0.62$\pm$0.11\\ 
Baade-Wesselink (RR Cet)& 0.55$\pm$0.12\\
HB stars: Evolutionary models - bright& 0.43$\pm$0.12\\
HB stars: Evolutionary models - faint& 0.56$\pm$0.12\\
Pulsation models (visual) & 0.58$\pm$0.12\\
Pulsation models ($PL_{K}Z$) & 0.59$\pm$0.10\\
Pulsation models (double-modes) & 0.57$\pm$0.06\\
Fourier parameters& 0.61$\pm$0.05\\
\hline
Main Sequence Fitting& 0.56$\pm$0.07\\
\hline
\end{tabular}
\label{t:tab7}
\end{table} 

\section{Ages}

\subsection{Derivation of ages}

Once distances and metallicities are known, ages can be derived from the
luminosity of the TO. Relevant values are listed in Table~\ref{t:tab3}. We
found the literature estimates of the apparent magnitude of the TO point
rather unsatisfactory for these three clusters. Therefore, we redetermined them
using a polynomial fit of the mean loci in the TO region. Parabolic fits were
used for NGC6397 and NGC6752, and a cubic fit for 47 Tuc. Again, we repeated
the procedure both for the $V,(B-V)$\ and the $V,(b-y)$\ diagrams, and
averaged the results. We attached to these average values errors in the range
0.02-0.05 mag, as estimated from the difference of the two determinations.
Absolute magnitudes of the TO of the three clusters were then derived by
simply combining our distance estimates with the apparent magnitude of the TO.
Finally, ages $t$\ were estimated by entering the values of the absolute
magnitude of the TO, along with the cluster metallicity corrected for the
$\alpha-$element overabundance, in the equation $\log
t=0.425~M_V-0.132~[M/H]-0.776$\ derived by Straniero et al. (1997). This
formula is based on standard isochrones, computed with no consideration of
sedimentation due to microscopic diffusion (see below). Note that the
Straniero et al. (1997) isochrones were computed assuming an helium content
of $Y=0.23$. Recent cosmological data rather suggests a slightly larger value
of $Y\sim 0.245$\ (see e.g. Spergel et al. 2003), in agreement with
determinations for globular clusters based on the so-called $R-$method
(Cassisi et al. 2003). Adopting such a higher helium content would result in
ages about 0.3 Gyr younger (see e.g. D'Antona et al. 2002). We then corrected
our age estimate for this effect, that is about 1/5 of our final error bar on
the absolute ages for the globular clusters.

The ages we determined for the three clusters are $13.9\pm 1.1$,
$13.8\pm 1.1$, and $11.2\pm 1.1$~Gyr for NGC6397, NGC6752, and 47 Tuc
respectively. The uncertainties were obtained by error propagation from the
errors in the distance derivations and in the metallicities, and do not take
into account the systematics of residual differences in the adopted
temperature scales for field and cluster stars (see Sect. 3.2), and of
adopted stellar models. The former contribute about 0.4~Gyr to the
uncertainty in ages. To estimate the contribution of the models, we used a
MonteCarlo approach similar to that described by Carretta et al. (2000). A
more rigorous version of this approach was presented by Chaboyer et al.
(1996), who obtained very similar results\footnote{The main difference
between the two approaches is that contribution by the various sources of
errors are assumed to be uncorrelated in the analysis of Carretta et al.,
while possible correlations are included in the approach of Chaboyer et al.}.
We considered several sources of theoretical uncertainties (treatment of
convection and microscopic diffusion, code, normalization to the Solar
absolute magnitude) and assumed a suitable distribution for the impact on
ages of each of these contributors. We then combined the individual errors in
random extractions, and estimated the standard deviation of the final
distribution. This value was then attached to the observed value as
representative of theoretical uncertainties. Shapes and limits of the adopted
distribution for each error source are given in Table~\ref{t:tab6}.

\begin{figure} 
\includegraphics[width=8.8cm]{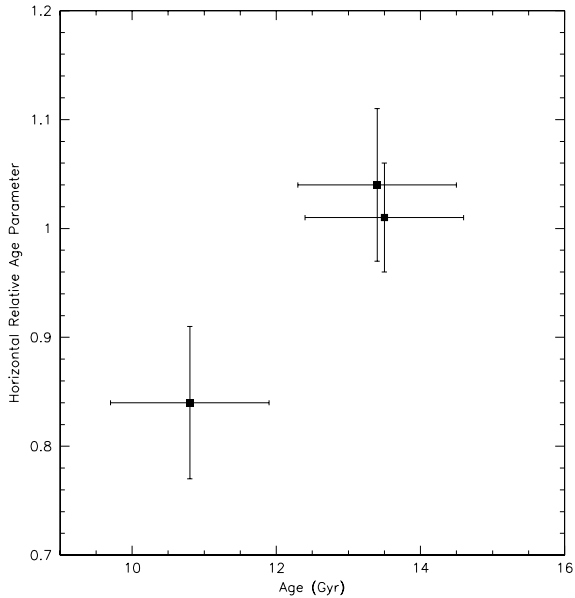}
\caption[]{Comparison between our absolute ages for the program clusters
and the relative age determined by Rosenberg et al.
(1999) with the horizontal method}
\label{f:fig3}
\end{figure}

\begin{table}
\caption{Model Error budget}
\begin{tabular}{lcc}
\hline
Error Source    & Distribution & Limits       \\
                &              & (Gyr)        \\
\hline
Convection      & Flat         & $-0.4$, 0.4  \\
Code            & Flat         & $-0.4$, 0.4  \\
Micr. Diffusion & Gaussian     & 0.4 at $-0.4$ \\
Solar $M_V$     & Flat         & $-0.3$, 0.3  \\
\hline
\end{tabular}
\label{t:tab6}
\end{table}

We note here that the impact of microscopic diffusion is strongly constrained
by our results for NGC6397 (Gratton et al. 2001): they clearly disagree with
results obtained by several authors, assuming complete ionization (see
Chaboyer et al. 2002). A careful discussion of this topic is given by Richard
et al. (2002), who presented results from more sophisticated models that
account for partial ionization and the effect of levitation due to radiation
pressure. However, also the predictions from these more advanced models do not
agree with the results of Gratton et al. (2001), unless it is arbitrarily
assumed that there is some mixing at the base of the outer convective envelope,
possibly induced by some turbulence. Once this is considered, the impact of
microscopic diffusion is constrained to be at most half of the value predicted
by models with complete ionization, that is not larger than about 0.5~Gyr. In
our estimates we have assumed that uncertainties in the treatment of
microscopic diffusion may be represented by a gaussian with a standard
deviation of $\pm 0.4$~Gyr, centered at a value of $-0.4$~Gyr, since some
small impact of microscopic diffusion is still expected, even if turbulence is
present (see Chaboyer et al. 2002, and VandenBerg et al. 2002). From this
approach we conclude that the theoretical error bar is 0.6~Gyr, and that a
downward correction of 0.4~Gyr should be applied to the ages derived from
isochrones that do not include sedimentation due to microscopic diffusion.

Our data suggests that 47 Tuc may be younger than NGC6397 and NGC6752. The age
difference (2.6~Gyr) is slightly larger than the sum of the error bars. A
similar age difference between these clusters was obtained by Rosenberg et al.
(1999) and Salaris \& Weiss (2002) from differential age determinations.
Fig.~\ref{f:fig3} compares relative and absolute age estimates for the three
clusters. The suggestion that 47 Tuc is indeed younger than the two other
clusters is strong, although perhaps not yet conclusive. It seems thus wiser
not to include 47 Tuc when determining the age of the oldest clusters in our
Galaxy. On the other hand, these same studies also provide strong evidences
that both NGC6397 and NGC6752 are nearly coeval to the oldest clusters in the
Galaxy. The average age for these two clusters (corrected downward by 0.4~Gyr
as discussed above) is then $13.4\pm 0.8\pm 0.6$~Gyr; we will assume
that this is the age of the oldest Galactic globular clusters. 47 Tuc (with
other disk clusters) is likely about 2.6~Gyr younger.

Our age estimate for the oldest globular clusters agrees well with the other
evaluations mentioned in the Introduction. It is nearly coincident with the
value of $14.0\pm 2.4$~Gyr determined in a completely independent way using
nucleocosmochronology for the extremely metal-poor star CS31082-001 by Hill et
al. (2002). It agrees also fairly well with the age determined for M4 by
Hansen et al. (2002). However, in view of the large uncertainties present in
both these determinations, this agreement may be fortuitous. Finally, our age
of the oldest globular clusters coincides with the determination of the age
of the Universe by the WMAP group (Spergel et al. 2003).

\begin{figure} 
\includegraphics[width=8.8cm]{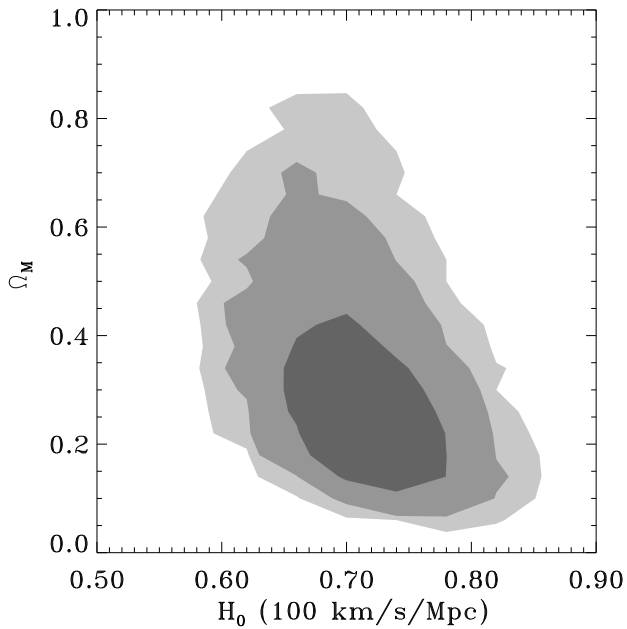}
\caption[]{Values of the matter density parameter $\Omega_M$\ as a function of
the Hubble constant $H_0$\ in a flat Universe, obtained from a 10,000 random
extraction of the values of the age of the Universe, assumed to be 0.3~Gyr
older than the age of the oldest globular clusters ($13.4\pm 1.4$~Gyr), and of
$H_0$ determined by Spergel et al. (2003) from WMAP experiment and by the HST
Key-Project (Freedman et al. 2001): $71\pm 4$~\kms\,Mpc$^{-1}$. Different gray
areas correspond to 67, 95 and 99\% levels of confidence}
\label{f:fig6}
\end{figure}

\begin{figure} 
\includegraphics[width=8.8cm]{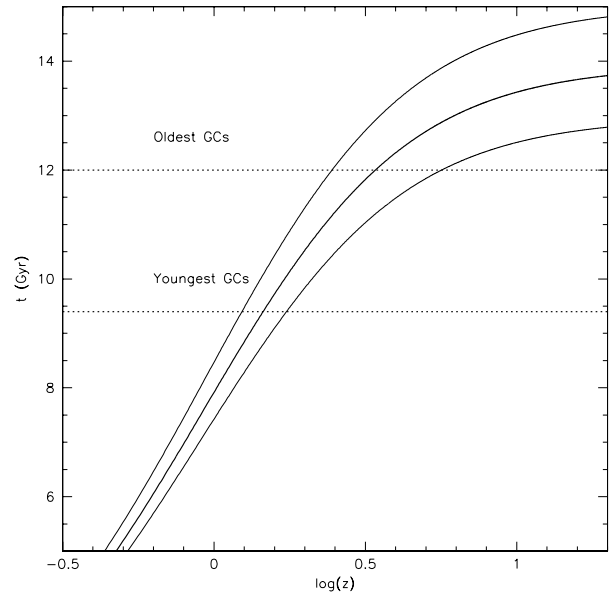}
\caption[]{Age as a function of redshift $z$\ for standard $\Lambda$CDM
Universe models. The thick solid line represents the curve obtained for the
best choice of parameters given by Spergel et al. (2003); the thin solid lines
are obtained summing 1-$\sigma$\ variations of the parameters (assumed to be
independent to each other). Dashed lines are the 1-$\sigma$\ lower limits to
the age of the oldest (NGC6397 and NGC6752) and youngest globular clusters (47
Tuc)}
\label{f:fig4}
\end{figure}

\subsection{The age of the Galaxy in a cosmological framework}

Our estimate of $13.4\pm 0.8\pm 0.6$~Gyr for the age of the oldest
globular clusters provides useful constraints for cosmology. Since the
Universe must be older than the globular clusters, its age is constrained to
be larger than 12.0~Gyr at 1-$\sigma$\ level, and larger than 10.6~Gyr 
at a 2-$\sigma$\ level. On the other hand, the WMAP data (Spergel
et al. 2003) and the HST key-project results (Freedman et al. 2001) constrain
the value of $H_0$\ at $71\pm 4$~\kms\,Mpc$^{-1}$\ (Spergel et al. 2003).
Assuming that the Universe is 0.3~Gyr older than the oldest globular
clusters, by performing integration of the appropriate equation using 10,000
values of the age of the Universe and of $H_0$\ randomly extracted around
these values (see Fig.~\ref{f:fig6}), we found that the value of
$\Omega_M$\ in a flat Universe (as indicated by the microwave background
data: see Spergel et al. 2003) is constrained to be $\Omega_M<0.57$\ and
$\Omega_M<0.75$\ at 95\% and 99\% level of confidence, respectively. This
result indicates the strong need for a dark energy even without taking into
account the results obtained from the high redshift Type Ia SNe (Perlmutter
et al. 1999), and those given by galaxy clusters (see e.g. the review by
Rosati et al. 2002).

An alternative use of our age estimate is to constrain the exponent $w$\ of
the equation of state for the dark energy, as described in Jimenez et al.
(2003). An adequate discussion requires the use of cosmological models; this is
out of the purposes of this paper. However, we note that our age
estimate is more precise and compelling than that considered by Jimenez et al.
(2003), so that more stringent lower limits to the time averaged value of $w$
may be derived by combining our age estimate with the location and shape of
the first acoustic peak in the microwave background spectrum (Jimenez et al.
2003).

\subsection{The epoch of Galaxy formation}

It is widely assumed that Galactic globular clusters probe the formation of
halo and thick disk of our Galaxy (Peebles \& Dicke 1968). This is supported
by their kinematics (Zinn 1985), chemical composition (Carney 1993), and
colours (Gilmore et al. 1995). However, it should be recalled that in some
external galaxies clusters are systematically bluer than the underlying field
population, indicating significant systematic differences (see e.g. Harris
1991). On the whole, it seems however reasonable to infer from the age of
globular clusters the epoch of formation of the earliest stellar populations
in our Galaxy. An earlier attempt in this sense was made by Carretta et al.
(2000), who concluded that Galactic globular clusters formed at $z>1$. A much
more refined evaluation is possible now, thanks to our new determinations of
the ages of NGC6397, NGC6752, and 47 Tuc, and the very recent and precise
estimates of the age and geometry  of the Universe from the WMAP experiment.
In a standard $\Lambda$CDM scenario (Bennett et al. 2003; Spergel et al. 2003)
according to WMAP the age of the Universe is $13.7\pm 0.2$~Gyr; the Hubble
constant (thanks also to data by the HST Key Project: Freedman et al. 2001) is
$H_0=71\pm 4$~km$\,$s$^{-1}\,$Mpc$^{-1}$; and the values of $\Omega_M$\ and
$\Omega_\Lambda$\ are fixed at $0.27\pm 0.04$\ and $0.73\pm 0.04$\
respectively.

When coupled with our age estimate for the oldest Galactic globular clusters
of $13.4\pm 0.8\pm 0.6$~Gyr, this leads to the conclusion that they
formed within 1.7~Gyr from the Big Bang at 1-$\sigma$\ level of confidence;
this corresponds to a redshift of $z>2.5$\ (see Fig.~\ref{f:fig4}).

Our age for 47 Tuc, as well as the relative age estimates by Rosenberg et al.
(1999) and Salaris \& Weiss (2002), indicate that the epoch of formation of
globular clusters lasted about 2.6~Gyr, ending about $10.8\pm 1.4$~Gyr ago,
that corresponds to a redshift of $z>1.3$. The reason for the absence of a
significant population of younger globular clusters in the Galaxy is not
entirely understood; however, it is likely to be connected with the presence
of a thin disk. In fact, several authors since Peebles \& Dicke (1968) related
the large overdensities required to the formation of globular clusters to
violent phases connected to accretion episodes, that would have disrupted or
at least considerably heated the thin disk. It is then interesting that a
phase of low star formation followed the formation of the thick disk (see
Gratton et al. 1996, 2000; Fuhrmann 1998; Liu \& Chaboyer 2000), and of its
population of globular clusters, like 47 Tuc (Zinn 1985), before the formation
of the thin disk started. After this phase, only a few globular clusters
actually formed; most of these objects (like e.g. Pal 12) are likely connected
to the Sagittarius dwarf galaxy (see e.g. Dinescu et al. 2000).

In the standard $\Lambda$CDM the end of the phase of formation of globular
clusters in our Galaxy corresponds to a redshift of $z>1.3$, and suggests a
close link between the epoch of formation of the Milky Way spheroid, and that
of the early spheroids in high redshift galaxies (Madau et al. 1998).

\begin{acknowledgements}
A special thanks goes to P. Montegriffo for making available to us his
software. We thank the anonymous referee for his/her useful suggestions
that helped to improve the paper. This research has made use of the SIMBAD
data base, operated at CDS, Strasbourg, France; it was funded by Ministero
Universit\`a e Ricerca Scientifica, Italy, through COFIN 2001028897.
\end{acknowledgements}

\end{document}